\pdfoutput=1
\documentclass[11pt,a4paper]{article}
\usepackage{jcappub}
\usepackage{booktabs}

\begin{document}

%%% BB
\newcommand{\bb}{\ensuremath{\beta\beta}}
%%% BB0NU
\newcommand{\bbonu}{\ensuremath{\beta\beta0\nu}}
%%% BB2NU
\newcommand{\bbtnu}{\ensuremath{\beta\beta2\nu}}

% NME
\newcommand{\Monu}{\ensuremath{\Big|M_{0\nu}\Big|}}
\newcommand{\Mtnu}{\ensuremath{\Big|M_{2\nu}\Big|}}
% PHASE-SPACE FACTOR
\newcommand{\Gonu}{\ensuremath{G_{0\nu}(\Qbb, Z)}}
\newcommand{\Gtnu}{\ensuremath{G_{2\nu}(\Qbb, Z)}}

% MBB
\newcommand{\mbb}{\ensuremath{m_{\beta\beta}}}
\newcommand{\kgy}{\ensuremath{\rm kg \cdot y}}
\newcommand{\ckky}{\ensuremath{\rm counts/(keV \cdot kg \cdot y)}}
\newcommand{\mbba}{\ensuremath{m_{\beta\beta}^a}}
\newcommand{\mbbb}{\ensuremath{m_{\beta\beta}^b}}
\newcommand{\mbbt}{\ensuremath{m_{\beta\beta}^t}}
\newcommand{\nbb}{\ensuremath{N_{\beta\beta^{0\nu}}}}

% Qbb
\newcommand{\Qbb}{\ensuremath{Q_{\beta\beta}}}

% Tonu
\newcommand{\Tonu}{\ensuremath{T_{1/2}^{0\nu}}}

% Tonu
\newcommand{\Ttnu}{\ensuremath{T_{1/2}^{2\nu}}}

%% Ca-48
\newcommand{\CA}{\ensuremath{^{48}\rm Ca}}
%% Ge-76
\newcommand{\GE}{\ensuremath{^{76}\rm Ge}}
%% Se-82
\newcommand{\SE}{\ensuremath{^{82}\rm Se}}
%% Mo-100
\newcommand{\MO}{\ensuremath{^{100}{\rm Mo}}}
%% Te-130
\newcommand{\TE}{\ensuremath{^{130}{\rm Te}}}
%% Xe-136
\newcommand{\XE}{\ensuremath{^{136}\rm Xe}}
%% Nd-150
\newcommand{\ND}{\ensuremath{^{150}{\rm Nd}}}

% Bi-214
\newcommand{\Bi}{\ensuremath{^{214}}Bi}
% Tl-208
\newcommand{\Tl}{\ensuremath{^{208}}Tl}
% Pb-208
\newcommand{\Pb}{\ensuremath{^{208}}Pb}
% Pb-208
\newcommand{\PBD}{\ensuremath{^{210}}Pb}

% Po-214
\newcommand{\Po}{\ensuremath{^{214}}Po}

% bru
\newcommand{\bru}{cts/(keV$\cdot$kg$\cdot$y)}

\newcommand{\ZR}{\ensuremath{{}^{96}{\rm Zr}}}
\newcommand{\KR}{\ensuremath{{}^{82}{\rm Kr}}}

\newcommand{\GES}{\ensuremath{{}^{68}\rm Ge}}
\newcommand{\TL}{\ensuremath{{}^{208}\rm{Tl}}}

\newcommand{\CO}{\ensuremath{{}^{60}\rm Co}}
\newcommand{\PO}{\ensuremath{{}^{214\rm Po}}}
\newcommand{\U}{\ensuremath{{}^{235}\rm U}}
\newcommand{\CT}{\ensuremath{{}^{10}\rm C}}
\newcommand{\BE}{\ensuremath{{}^{11}\rm Be}}
\newcommand{\BO}{\ensuremath{{}^{8}\rm Be}}
\newcommand{\BA}{\ensuremath{{}^{136}\rm Ba}}
\newcommand{\UDTO}{\ensuremath{{}^{238}\rm U}}
\newcommand{\CD}{\ensuremath{^{116}{\rm Cd}}}
\newcommand{\THO}{\ensuremath{{}^{232}{\rm Th}}}
\newcommand{\BI}{\ensuremath{{}^{214}}Bi}
\newcommand{\RN}{\ensuremath{{}^{222}}Rn} %% Load some useful commands

%%%%%%%%%%%%%%%%%%%%%%%%%%%%%%%%%%%%%%%%%%%%%%%%%%%%%%%%%%%%

\title{GraXe, graphene and xenon for neutrinoless double beta decay searches}

\author[a]{J.J.~G\'omez-Cadenas,} \emailAdd{gomez@mail.cern.ch}
\author[b]{F.~Guinea,} \emailAdd{paco.guinea@icmm.csic.es}
\author[c]{M.M.~Fogler,}\emailAdd{mfogler@ucsd.edu}
\author[d]{M.I.~Katsnelson,}\emailAdd{katsnel@sci.kun.nl}
\author[a]{J.~Mart\'in-Albo,} \emailAdd{justo.martin-albo@ific.uv.es}
\author[a]{F.~Monrabal,} \emailAdd{francesc.monrabal@ific.uv.es}
\author[a]{J.~Mu\~noz Vidal} \emailAdd{jmunoz@ific.uv.es}

\affiliation[a]{Instituto de F\'isica Corpuscular (IFIC), CSIC \& Universitat de Valencia \\ Calle Catedr\'atico Jos\'e Beltr\'an, 2, 46980 Valencia, Spain}
\affiliation[b]{Instituto de Ciencia de Materiales de Madrid (ICMM), CSIC\\
Calle Sor Juana In\'es de la Cruz, 3, 28049 Madrid, Spain}
\affiliation[c]{Department of Physics, University of California San Diego\\
9500 Gilman Drive, La Jolla, California 92093, USA}
\affiliation[d]{Institute for Molecules and Materials, Radboud University Nijmegen\\Heijendaalseweg 135, 6525 AJ Nijmegen, The Netherlands}

%%%%%%%%%%%%%%%%%%%%%%%%%%%%%%%%%%%%%%%%%%%%%%%%%%%%%%%%%%%%
\abstract{
We propose a new detector concept, \emph{GraXe} (to be pronounced as \emph{grace}), to search for neutrinoless double beta decay in \XE. GraXe combines a popular detection medium in rare-event searches, \emph{liquid xenon}, with a new, background-free material, \emph{graphene}. 

In our baseline design of GraXe, a sphere made of graphene-coated titanium mesh and filled with liquid xenon (LXe) enriched in the \XE\ isotope is immersed in a large volume of natural LXe instrumented with photodetectors. Liquid xenon is an excellent scintillator, reasonably transparent to its own light. Graphene is transparent over a large frequency range, and impermeable to the xenon. Event position could be deduced from the light pattern detected in the photosensors. External backgrounds would be shielded by the buffer of natural LXe, leaving the ultra-radiopure internal volume virtually free of background. 

Industrial graphene can be manufactured at a competitive cost to produce the sphere. Enriching xenon in the isotope \XE\ is easy and relatively cheap, and there is already near one ton of enriched xenon available in the world (currently being used by the EXO, KamLAND-Zen and NEXT experiments). All the cryogenic know-how is readily available from the numerous experiments using liquid xenon. An experiment using the GraXe concept appears realistic and affordable in a short time scale, and its physics potential is enormous.}

\arxivnumber{1110.6133}

\maketitle

%%%%%%%%%%%%%%%%%%%%%%%%%%%%%%%%%%%%%%%%%%%%%%%%%%%%%%%%%%%%

\section{Introduction} \label{sec:intro}
%%%
Neutrinoless double beta decay (\bbonu) is a hypothetical, very slow nuclear transition in which two neutrons undergo $\beta$-decay simultaneously and without the emission of neutrinos. The importance of this process goes beyond its intrinsic interest: an unambiguous observation would establish that neutrinos are Majorana particles --- that is to say, truly neutral particles identical to their antiparticles --- and prove that total lepton number is not a conserved quantity. 

After 70 years of experimental effort, no compelling evidence for the existence of \bbonu\ has been obtained. However, a new generation of experiments that are already running or about to run promises to push forward the current limits exploring the \emph{degenerate} region of neutrino masses (see \cite{GomezCadenas:2011it} for a recent review of the field). In order to do that, the experiments are using masses of \bb\ isotope ranging from tens of kilograms to several hundreds, and will need to improve the background rates achieved by previous experiments by, at least, an order of magnitude. If no signal is found, masses in the ton scale and further background reduction will be required. Among the new-generation experiments, only a few can possibly be extrapolated to those levels.

In this paper, we propose a new detector concept that can result in an experiment possessing both a very large isotope mass and an extremely low background rate. We call this detector \emph{GraXe} (to be pronounced as \emph{grace}), contracting the two keywords that define our proposal: \emph{graphene} and \emph{xenon}. 

GraXe combines several ideas already being exploited in the field with new possibilities available thanks to the use of graphene. In the simplest version of the detector, a balloon of graphene filled with liquid xenon (LXe) enriched in the isotope \XE, a \bb\ emitter, would be placed in the center of a large LXe scintillation detector, such as future versions of XMASS \cite{Suzuki:2000qy,Abe:2010zz}. Possible improvements of this baseline include adding an electrode to the center of the detector to measure the ionization, therefore improving the location of the event and possibly the energy resolution.

Liquid xenon offers as a detection medium high stopping power (thus the capability of shielding easily external backgrounds), excellent radiopurity and the availability of ionization and scintillation signals. Furthermore, it is possible to deploy a large mass of enriched xenon, being the simplest (and cheapest) \bb\ source to enrich. In addition, GraXe exploits the fact that graphene is: impermeable to the xenon (no diffusion losses) and with enormous tensile strength; transparent to the VUV light emitted by xenon; metallic; and extremely radiopure (virtually \emph{zero} contamination of radioactive impurities) and non-degassing (thus the LXe contained in the graphene balloon is essentially free of impurities). 

The paper is organized as follows. Section \ref{sec:bb0nu} briefly discusses the physics motivations to search for \bbonu\ processes, as well as the basic experimental aspects. Section \ref{sec:LXe} reviews the fundamental properties of LXe as a detection medium, and section \ref{sec:graphene} offers a very condensed summary of the properties of graphene of interest for our application. The conceptual ideas behind GraXe are developed in sections \ref{sec:sci} and \ref{sec:ion}. Finally, section \ref{sec:map} presents conclusions and an outlook, including a possible road map to develop the experiment.

\section{Neutrinoless double beta decay} \label{sec:bb0nu}
%%%%%%%%%%%%%%%%%%%%%%%%%%%%%%%%%%%%%%%%%%%%%%%%%%%%%%%%%%
\subsection{Double beta decay and Majorana neutrinos}
Double beta decay (\bb) is a very rare nuclear transition in which a nucleus with $Z$ protons decays into a nucleus with $Z+2$ protons and the same mass number $A$. The decay can occur only if the initial nucleus is less bound than the final nucleus, and both more than the intermediate one. There are 35 naturally-occurring isotopes that can undergo \bb. Two decay modes are usually considered:
\begin{itemize}
\item The standard two-neutrino mode (\bbtnu), consisting in two simultaneous beta decays, $(Z,A) \rightarrow (Z+2,A) + 2\ e^{-} + 2\ \overline{\nu}_{e}$, which has been observed in several isotopes with typical half-lives in the range of $10^{18}$--$10^{21}$ years (see, for instance, \cite{GomezCadenas:2011it} and references therein). 
\item The neutrinoless mode (\bbonu), $(Z,A) \rightarrow (Z+2,A) + 2\ e^{-}$, which violates lepton-number conservation, and is therefore forbidden in the Standard Model of particle physics. An observation of \bbonu\ would prove that neutrinos are massive, Majorana particles \cite{Schechter:1981bd}. No convincing experimental evidence of the decay exists to date (see section \ref{subsec:bbexp}).
\end{itemize}

The implications of experimentally establishing the existence of \bbonu\ would be profound. First, it would demonstrate that total lepton number is violated in physical phenomena, an observation that could be linked to the cosmic asymmetry between matter and antimatter through the process known as \emph{leptogenesis} \cite{Fukugita:1986hr, Davidson:2008bu}. Second, Majorana neutrinos provide a natural explanation to the smallness of neutrino masses, the so-called \emph{seesaw mechanism} \cite{Minkowski:1977sc, GellMann:1980vs,Yanagida:1979, Mohapatra:1979ia}.

Several underlying mechanisms --- involving, in general, physics beyond the Standard Model  --- have been proposed for \bbonu\ \cite{Rodejohann:2011mu}, the simplest one being the virtual exchange of light Majorana neutrinos. Assuming this to be the dominant one at low energies, the half-life of \bbonu\ can be written as
\begin{equation}
(T^{0\nu}_{1/2})^{-1} = G^{0\nu} \ \big|M^{0\nu}\big|^{2} \ \mbb^{2}.
\end{equation}
In this equation, $G^{0\nu}$ is an exactly-calculable phase-space integral for the emission of two electrons; $M^{0\nu}$ is the nuclear matrix element of the transition, that has to be evaluated theoretically; and \mbb\ is the \emph{effective Majorana mass} of the electron neutrino:
\begin{equation}
\mbb = \Big| \sum_{i} U^{2}_{ei} \ m_{i} \Big| \ ,
\end{equation}
where $m_{i}$ are the neutrino mass eigenstates and $U_{ei}$ are elements of the neutrino mixing matrix. Therefore, a measurement of the \bbonu\ decay rate would provide direct information on neutrino masses \cite{GomezCadenas:2011it}.

%%%%%%%%%%%%%%%%%%%%%%%%%%%%%%%%%%%%%%%%%%%%%%%%%%%%%%%%%%
\subsection{Experimental aspects} \label{subsec:bbexp}
\begin{figure}
\centering
\includegraphics[height=7cm]{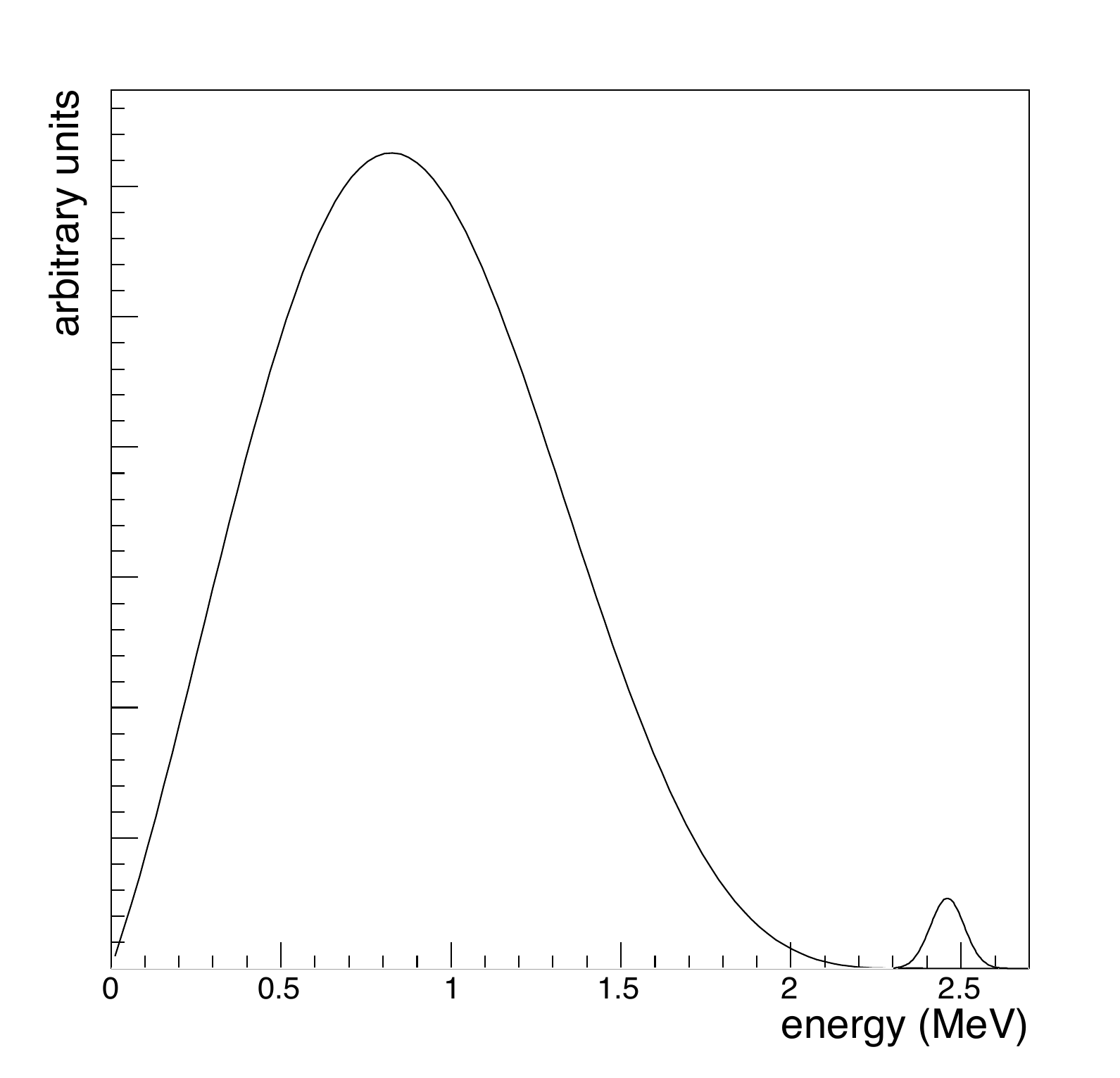}
\caption{Energy spectrum of the electrons emitted in the \bb\ decay of \XE, as seen with a 1\% FWHM energy resolution at \Qbb. The left peak corresponds to the \bbtnu\ decay, while the right peak, centered at $\Qbb = 2458$ keV, corresponds to the \bbonu. The normalization scale between the two peaks is arbitrary.} \label{fig:bbspectrum}
\end{figure}

The detectors used in double beta decay experiments are designed to measure the energy of the radiation emitted by a \bb\ source. In the case of \bbonu, the sum of the kinetic energies of the two released electrons is always the same, and corresponds to the mass difference between the parent and the daughter nuclei: $\Qbb \equiv M(Z,A)-M(Z+2,A)$. However, due to the finite energy resolution of any detector, \bbonu\ events are reconstructed within a non-zero energy range centered around \Qbb, typically following a gaussian distribution, as shown in figure \ref{fig:bbspectrum}. Other processes occurring in the detector can fall in that region of energies, thus becoming a background and compromising drastically the experiment's expected sensitivity to \mbb\ \cite{GomezCadenas:2010gs}.

%%%%%
%%%%%

All double beta decay experiments have to deal with an intrinsic background, the \bbtnu, that can only be suppressed by means of good energy resolution. Backgrounds of cosmogenic origin force the underground operation of the detectors. Natural radioactivity emanating from the detector materials and surroundings can easily overwhelm the signal peak, and consequently careful selection of radiopure materials is essential. Additional experimental signatures that allow the distinction of signal and background are a bonus to provide a robust result.

The Heidelberg-Moscow experiment set the most sensitive limit to the half-life of \bbonu\ so far: $T^{0\nu}_{1/2}(\GE) \ge 1.9\times10^{25}$ years \cite{KlapdorKleingrothaus:2000sn}. In addition, a subgroup of the experiment observed evidence of a positive signal, with a best value for the half-life of $1.5\times10^{25}$ years \cite{KlapdorKleingrothaus:2001ke}, corresponding to a Majorana neutrino mass of about 0.4 eV. The claim was very controversial \cite{Aalseth:2002dt}, and still awaits an experimental response. A new generation of \bb\ experiments --- already running or about to do so --- will push the current limits down to neutrino masses of about 100 meV or better \cite{GomezCadenas:2011it}.

\section{Liquid xenon as detection medium} \label{sec:LXe}
%%%
Liquid xenon combines several physical properties that make it an attractive detection medium: it has the highest stopping power among the liquid noble elements, thanks to its high atomic number ($Z=54$) and its density (3.1 g/cm$^{3}$); it provides both an ionization and a scintillation signal, the latter being comparable in intensity to that of NaI and with a faster time response; being a noble element, it is inert and easy to purify; and it is reasonable abundant and not too expensive, allowing the construction of large detectors.

Two naturally-occurring isotopes of xenon can decay \bb, $^{134}$Xe ($\Qbb = 825$ keV) and $^{136}$Xe ($\Qbb = 2458$ keV). The latter, having a higher $Q$-value, is preferred for neutrinoless double beta decay searches, because the decay rate is proportional to $\Qbb^{5}$ and the radioactive backgrounds are less abundant at higher energies. Besides, the \bbtnu\ mode of \XE\ is slow ($2.11\times10^{21}$ years \cite{Ackerman:2011gz}), and hence the experimental requirement for good energy resolution is less stringent than for other \bb\ sources. \XE\ constitutes 8.86\% of all natural xenon, but the enrichment process is relatively simple and cheap compared to that of other \bb\ isotopes. The detection properties of enriched xenon are equivalent to those of natural xenon.

%%%%%%%%%%%%%%%%%%%%%%%%%%%%%%%%%%%%%%%%%%%%%%%%%%%%%%%%%%
\subsection{Primary signals in liquid xenon: ionization and scintillation}
Charged particles interacting with liquid xenon lose their energy through two atomic processes: \emph{excitation}, where energy is transferred to an atomic electron that moves then to a higher energy state; and \emph{ionization}, which results in the formation of pairs of positively charged ions and free electrons. Both atomic de-excitations and \emph{recombination} of the ionization pairs lead eventually to the emission of \emph{scintillation} light of characteristic properties.

The average energy required to produce an ionization pair in liquid xenon is \cite{Aprile:2009dv}
\begin{equation}
\label{eq:Wi}
W_{\rm i} = 15.6 \pm 0.3 \ \mathrm{eV}\, .
\end{equation}
Since this quantity does not depend very strongly on the type and energy of the considered particle, the number of ionization charges can be used as a measure of the deposited energy.

The scintillation mechanism of liquid xenon is well understood \cite{Aprile:2009dv}. The emission spectrum extends from the infrared to the vacuum ultraviolet (VUV), where it peaks at $\sim 178$ nm. The scintillation yield (i.e., the number of emitted photons) depends on the linear energy transfer (LET) of a particle, that is, the density of ionization pairs produced along the track, and for that reason, it depends on the type of particle. For relativistic electrons, the average deposited energy in liquid xenon required to create one scintillation photon is \cite{Aprile:2009dv}
\begin{equation}
\label{eq:Ws}
W_{\rm s} = 21.6~\mathrm{eV}\, .
\end{equation}

Both signals, scintillation and ionization, can be observed in LXe, and their amplitudes are strongly anti-correlated \cite{Aprile:2009dv}.

%%%%%%%%%%%%%%%%%%%%%%%%%%%%%%%%%%%%%%%%%%%%%%%%%%%%%%%%%%
\subsection{Collection of the ionization charges}
Detection of the ionization signal generally implies the so-called \emph{drift} of the charge carriers (electrons and/or ions) under the influence of an external electric field. At low fields, the electron drift velocity, $v_{\rm d}$, is almost proportional to the field strength, $E$, with the electron mobility, $\mu$, as the proportionality constant: $v_{\rm d} = \mu E$. In liquid xenon, the electron mobility is about \mbox{2000 cm$^2$ V$^{-1}$ s$^{-1}$} \cite{Aprile:2009dv}. At high fields, the electron drift velocity saturates, becoming independent of the electric field intensity. The mobility of the positive carriers is several orders of magnitude smaller than electron mobility, about \mbox{$4\times10^{-3}$ cm$^2$ V$^{-1}$ s$^{-1}$} \cite{Aprile:2009dv}.

Charge carriers deviate from the drift lines defined by the electric field due to \emph{diffusion}, limiting the position resolution of the detector. The rms spread in the transversal (longitudinal) direction of drift is given by
\begin{equation}
\sigma_{\rm T(L)} = \sqrt{D_{\rm T(L)}\ t}\ ,
\end{equation}
where $D_{T(L)}$ is the transversal (longitudinal) \emph{diffusion coefficient} and $t$ is the drift time. For electric field strengths in the range 1--10 kV/cm, $D_{\rm T}\simeq100$ cm$^{2}$/s. The longitudinal coefficient is about $1/10$ of the transverse coefficient, contributing a negligible amount to the position resolution. 

Electron attachment to electronegative impurities dissolved in the LXe may lead to a significant decrease of the ionization signal. The concentration of impurities must be kept under control (typically below 1 ppb) recirculating the xenon through the appropriate filters.

%%%%%%%%%%%%%%%%%%%%%%%%%%%%%%%%%%%%%%%%%%%%%%%%%%%%%%%%%%
\subsection{Detection of scintillation light}
The VUV emission spectrum of LXe is still accessible for photomultipliers equipped with VUV-graded windows, allowing direct detection of the scintillation photons. Knowledge of the optical properties of LXe in the VUV is essential to understand the performance of the detector.

The transparency of liquid xenon to its own scintillation light its limited by Rayleigh scattering and the possible presence of dissolved impurities. The light attenuation can be described by a negative exponential:
\begin{equation}
I(x) = I_{0}\ e^{-x/\lambda_{\rm att}} \label{eq:att},
\end{equation}
where $\lambda_{\rm att}$ is the photon attenuation length, which consists of two separate components: 
\begin{equation}
\frac{1}{\lambda_{\rm att}} = \frac{1}{\lambda_{\rm abs}} + \frac{1}{\lambda_{\rm sca}}. \label{eq:latt}
\end{equation}
The absorption length, $\lambda_{\rm abs}$, describes true absorption and loss of photons by impurities, and the scattering length, $\lambda_{\rm sca}$, represents
elastic scattering of photons without any loss. The latter is dominated by Rayleigh scattering and estimated to be about 30 cm \cite{Seidel:2001vf, Baldini:2004ph}. Absorption lengths longer than 100 cm can be achieved using suitable purification techniques \cite{Baldini:2004ph}. Since the scattered photons are not totally lost, the effective attenuation length is longer than the value obtained substituting the previous numbers in equation (\ref{eq:latt}) \cite{Baldini:2004ph, Ueshima:2010}.

Knowledge of the refractive index of LXe in the VUV region is relevant for the optimization of the light collection. Measurements range from 1.54 to 1.69 at 178 nm \cite{Aprile:2009dv}.

%%%%%%%%%%%%%%%%%%%%%%%%%%%%%%%%%%%%%%%%%%%%%%%%%%%%%%%%%%
\subsection{Energy resolution in LXe}
Fluctuations in the number of electron-ion pairs produced by an ionizing particle limit the energy resolution that can be achieved in any detection medium. In 1947, U.~Fano demonstrated \cite{Fano:1947zz} that the variance in the number of charge carriers, $N$, produced by ionizing radiation is not given by Poisson statistics but by
\begin{equation}
\sigma^{2} = F N, \label{eq:fano}
\end{equation}
where the number $F$, known as \emph{Fano factor}, depends on the stopping material. 

Therefore, the ultimate energy resolution, often called the Fano-limit, achievable with a LXe ionization detector would be given by
\begin{equation}
\Delta E = 2.35 \ W_{\rm i} \ \sqrt{F_{\rm LXe}\ N} = 2.35\ \sqrt{F_{\rm LXe} \ W_{\rm i} \ E}\ ,
\end{equation}
where $\Delta E$ is the energy resolution expressed as a gaussian FWHM and $E$ is the energy of the ionizing radiation. The calculation of the Fano factor for LXe and other liquid rare gases was carried out by T.~Doke \cite{Doke:1981}, obtaining $F_{\rm LXe} = 0.059$. However, the corresponding energy resolution has not been achieved experimentally. The best resolution measured with a LXe ionization chamber is 30 keV for $\gamma$-rays of 554 keV, at a very high field of 17 kV/cm \cite{Takahashi:1975zz}. This corresponds to a relative energy resolution of $\Delta E/E= 5.4$\%, more than four times worse, in fact, than the Poisson limit ($F=1$). The reasons of the discrepancy between the experimental and theoretical energy resolution of liquid xenon remain unclear; see \cite{Aprile:2009dv, Nygren:2009zz} and references therein for a detailed discussion. 

Although the scintillation yield of LXe is comparable (or even higher, for some types of particles) to the number of ionization pairs, the energy resolution achievable by measuring only the scintillation signal is in practice much worse. Finite geometric coverage and finite quantum efficiency of photodetectors lead to very small optical detection efficiencies.

Nevertheless, the combined measurement of both scintillation and ionization signals reduces the fluctuation in the summed signal to a lower level than that in each individual signal, resulting in a better energy resolution \cite{Aprile:2009dv, Conti:2003av}. 

\section{Graphene} \label{sec:graphene}
Graphene is carbon membrane one atom thick \cite{CastroNeto:2009zz}. It was initially obtained from exfoliating graphite crystals \cite{Netal04, Netal05}.  New fabrication methods that make use of carbon films deposited on metal surfaces have made possible the obtention of graphene samples of more than 1 cm in size \cite{Ketal09, Letal09, Betal10}. These samples are made of crystalline grains separated by grain boundaries \cite{Hetal11, Yetal11, Cetal11, Ketal11}. Graphene is metallic, and the carrier concentration can be tuned by a metallic gate. Single layer graphene is impermeable to all elements \cite{Betal08}.

Graphene is transparent over a large frequency range, from the infrared to the ultraviolet \cite{BSHF10}. The transparency of graphene is $1 - \pi \alpha$, where $\alpha \approx 1 / 137$ is the vacuum fine structure constant \cite{Netal08}. The opacity of a sample with $N$ layers is $N \times \pi \times \alpha$. 

Graphene has the largest Young's modulus (normalized to its thickness) of any material, $E \approx 340$ N/m \cite{LVKH08}. The breaking strength of graphene is $\sim42$ N/m, which corresponds to a maximum strain $\varepsilon_c \approx 12 \%$. This value is reduced by about one order of magnitude in polycrystalline samples \cite{Hetal11}. Graphene is an attractive material as sample holder \cite{Netal10}, and it has been proposed that single layer graphene can support macroscopic objects despite its low weight \cite{hammock}. The elastic properties of suspended graphene membranes also lead to interesting electronic properties \cite{FGK08}. 

The coupling between graphene and a substrate is material-dependent and not fully understood. The order of magnitude of the van der Waals interaction between graphene and most substrates is expected to be in the range $10^{-3}$--$10^{-2}$ eV \AA$^{-2}$ \cite{Sabio:2008}. These estimates are in good agreement with recent experimental results \cite{Koenig:2011}.

\section{GraXe in scintillation mode} \label{sec:sci}
\subsection{Description of the detector}
\begin{figure}[t!]
\centering
\includegraphics[width=0.75\textwidth]{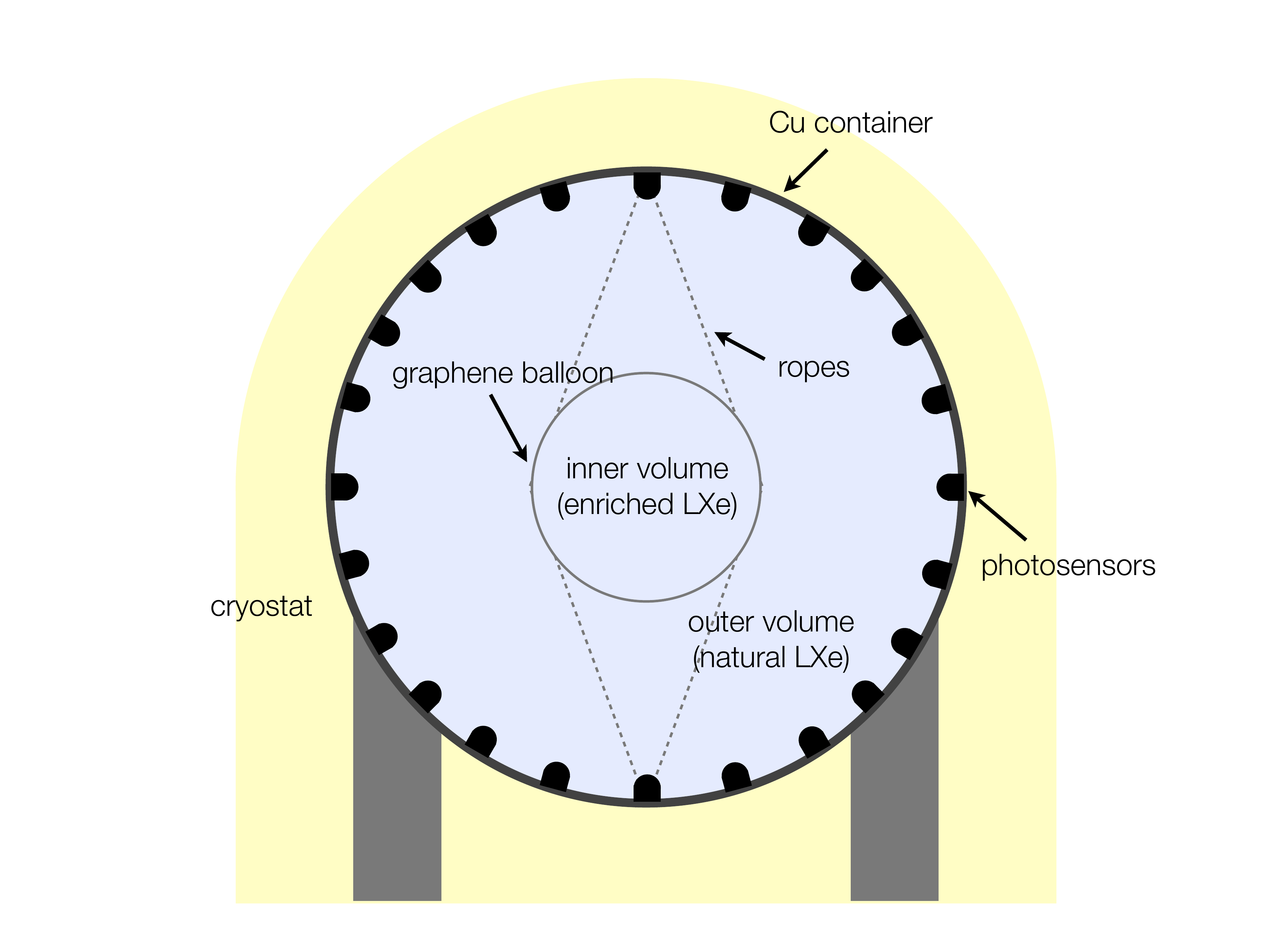}
\caption{In the simplest version of GraXe, a graphene balloon, $\sim45$ cm in radius and filled with \mbox{1 ton} of \XE-enriched liquid xenon, is fixed in the center of a large LXe scintillation detector.} \label{fig:GraXe}
\end{figure}

In its simplest version, GraXe would be a liquid xenon scintillation calorimeter. A conceptual sketch of the detector is shown in figure \ref{fig:GraXe}. A spherical container, 120 cm in radius and made of ultra-pure copper, would hold about 20 tons of LXe. In its center, a second sphere, the graphene balloon, $\sim45$ cm radius, would be fixed using low-background synthetic ropes and filled with 1 ton of LXe enriched in the isotope \XE. The container would be instrumented with an array of large photodetectors sensitive to the scintillation light of xenon. Event position reconstruction with a vertex resolution of a few centimeters can be deduced from the light pattern detected in the PMTs \cite{Ueshima:2010}. The liquid xenon outside the graphene balloon would act as a shield against external backgrounds, attenuating high-energy gammas by about 5 orders of magnitude, and leaving the very radiopure inner sphere virtually free of background. The entire detector could be surrounded by hydrocarbon material or water for additional shielding against neutrons \cite{Carson:2004cb}.

%%%%%
%%%%%

Even though the graphene balloon would be in hydrostatic equilibrium, it would need to withstand small pressure differences between the inside and the outside due to shock waves, local changes in density, etc. The strain developed by a balloon of radius $R$, under pressure $P$ and made of a material with two-dimensional Young's modulus $E$ is
\begin{equation}
\varepsilon \sim \frac{P\ R}{E}. \label{eq:strain}
\end{equation}
Assuming a safe value of 0.1 bar for the pressure difference, we find that $\varepsilon \simeq 1\, 300\%$ for a graphene balloon of radius $R=0.45$ m, well above the breaking threshold of single-layer graphene. The solution would be to deposit (or directly produce) the graphene layers on a metallic mesh, made of titanium, for example. The hole size in this mesh can be of the order of 100--500 $\mu$m diameter. The relation between strain and pressure for each hole is given approximately by equation (\ref{eq:strain}), with $R$ replaced by the radius of the hole, and hence $\varepsilon<1\%$. Therefore, single-layer graphene can be enough to make the container impermeable. 
For an interaction in the range considered here, graphene should adjust smoothly to the corrugations of the substrate, provided that the curvature radius is larger than 10--100 nm \cite{Kusminskiy:2011}. The transparency of the balloon is given by the open area of the mesh ($\sim90\%$) times the transparency of mono-layer graphene ($\sim98\%$), thus at least $85\%$. Its radioactivity can be negligible, due to its low mass (less than 100 g). If made with radiopure titanium \cite{Akerib:2011rr}, the total activity of the mesh would be less than 10 $\mu$Bq.\footnote{We consider only the thorium and uranium series, the natural decay chains relevant for this experiment. See section \ref{subsec:sensi_sci} for further details.} The graphene balloon would include an inlet tube for the LXe filling and recirculation. Notice, however, that there are no contaminants in contact with the enriched LXe, since the degassing of the balloon would be insignificant.

The detector container would be made of radiopure copper, like in the EXO \cite{Ackerman:2011gz} and XMASS experiments. Electro-formed copper has very low activity, about 5--10 $\mu$Bq/kg. Assuming a shell thickness of 2 cm, the total activity of the container would be about 35 mBq.

An obvious candidate photomultiplier for GraXe is the Hamamatsu R11410, a 3-inches tube specifically designed for radiopure operation in liquid xenon. It has a quantum efficiency (QE) of $\sim 26$\% at 175 nm, and a specific activity lower than $5$ mBq in each one of the relevant radioactive chains. Another candidate is the QUPID \cite{Teymourian:2011rs}, a new low background photosensor based on the hybrid avalanche photo-diode, and entirely made of ultra-clean synthetic fused silica. The QUPID has a diameter of 3 inches, 33\% quantum efficiency at 175 nm, and an activity of about 0.5 mBq in the U and Th series. If 100\% of the container is covered by 3-inches photodetectors, $\sim 4\,000$ will be needed. For a 70\% coverage, as in the XMASS detector, we will need $\sim2\,800$ devices. In this case and if QUPIDs are used, the overall activity of the photosensors array would be 2.8 Bq, dominating the radioactive budget of the detector.

%%%%%%%%%%%%%%%%%%%%%%%%%%%%%%%%%%%%%%%%%%%%%%%%%%%%%%%%%%%%%%%%%%%%
\subsection{Optical detection efficiency and energy resolution} \label{subsec:reso}
As explained in section \ref{sec:LXe}, an event of energy $E$ will produce $N_{\rm s} = E/W_{\rm s}$ scintillation photons in liquid xenon. Using the $Q$-value of the $\XE \rightarrow \BA$ transition, $\Qbb=2457.83$ keV \cite{Redshaw:2007un}, we obtain that
\begin{equation}
N_{\rm s} = 113\,788\ \mathrm{photons}\label{eq:Nsci}
\end{equation}
would be emitted in a \bbonu\ event.

Only a fraction of these photons will reach the surface of the photodetectors due to the partial opacity of the graphene balloon, $T$, the optical attenuation of LXe --- described by equation (\ref{eq:latt}) --- and the finite geometrical coverage, $G$. Also, the sensors themselves have a limited detection efficiency given by their light collection efficiency, $L_{c}$, times the quantum efficiency of their photocathode, $\eta$. Therefore, the number of photons \emph{actually} detected --- in other words, the number of photoelectrons (pe) emitted from the photocathodes of the sensors --- is
\begin{equation}
N_{\rm pe} = N_{\rm s} \cdot T \cdot G \cdot L_{c} \cdot \eta \cdot e^{x/\lambda_{\rm att}}.
\end{equation}

Since the process is dominated by Poisson statistics, one should expect an energy resolution (FWHM)
\begin{equation}
\Delta E/E  = \frac{2.35}{\sqrt{N_{\rm pe}}}\ .
\end{equation}

For a \bbonu\ event occurring in the center of GraXe, $x=120$ cm, and taking $\lambda_{\rm att}=50$ cm \cite{Baldini:2004ph}, $T=0.85$, $G=0.7$ and $L_{c}\cdot\eta=0.3$, we obtain that
\begin{equation}
N_{\rm pe} = 1843,
\end{equation}
resulting in an energy resolution of
\begin{equation}
\Delta E/E = 5.5\% \ \mathrm{FWHM}.
\end{equation}

The minimum observable energy corresponds to the level of the photodetectors' dark current. A conservative threshold of 50 keV can be assumed.

%%%%%%%%%%%%%%%%%%%%%%%%%%%%%%%%%%%%%%%%%%%%%%%%%%%%%%%%%%%%%%%%%%%%
\subsection{Sensitivity of GraXe in scintillation mode} \label{subsec:sensi_sci}
In order to gain a quantitative understanding of the performance of GraXe in scintillation mode we have written a Geant4 \cite{Agostinelli:2002hh} simulation of the detector. Four different types of events were generated: the signal, \bbonu, and the three backgrounds estimated to be the dominant, \bbtnu, \Tl\ and \Bi.\footnote{See section \ref{subsec:SensiIoni} for a comment on other sources of background.} The last two are radioactive by-products of the uranium and thorium series, respectively, and can be found as impurities in all materials. They are beta isotopes whose decay is followed by the emission of high-energy, de-excitation gammas. In particular, the $Q$-value of \XE\ is in the region between the photoelectric peaks of two of these gammas, at 2448 keV (from \Bi) and at 2615 keV (from \Tl). Other background sources such as muons, neutrons or neutrinos are estimated

Both \bbonu\ and \bbtnu\ were generated uniformly in GraXe's \emph{inner volume} (IV) --- that is, the space within the graphene balloon ---, while \Bi\ and \Tl\ were generated as emanating from the container's surface, representing the background from the container and from the photodetectors. The contribution of the balloon can be neglected at first order. A conservative energy resolution of 10\% FWHM at \Qbb\ was assumed in the simulation, about a factor of 2 worse than expected by photoelectron statistics (see section \ref{subsec:reso}). Only energy depositions larger than 50 keV are considered visible. 
Energy depositions separated by at least 6 degrees in the polar angle and 3 cm in radius are considered different \emph{clusters} (extrapolated from results of XMASS \cite{Ueshima:2010}). Single-cluster energy depositions within 1 FWHM around the $Q$-value are selected as signal-candidate events. 

%%%%%
\begin{table}[t!]
\begin{center}
\begin{tabular}{lcccc}
\toprule
Cuts & \bbonu\ & \bbtnu\ & \BI\ &  \TL\ \\ \midrule
Initial sample & $1.0\times10^5$ & $1.1\times10^9$ & $10^9$ & $10^8$ \\ 
No energy in OV & $9.6\times10^4$ & $1.9\times10^6$ & 52 & 50 \\ 
Only one cluster & $8.9\times10^4$ & $1.8\times10^6$ & 21 & 16 \\
ROI (1 FWHM) & $6.7\times10^4$ & $2.3\times10^4$ & 5 & 6 \\ \midrule
%
%Right side ROI  & $3. 2 \times 10^4$ & $1.5 \times 10^3$ & $2.0 $ & $6.0$\\ \hline

Rejection factor  & $0.67$ & $2.0 \times 10^{-5}$ & $5.0 \times 10^{-9}$ & $6.0\times 10^{-8}$ \\ 
\bottomrule
\end{tabular}
\end{center}
\caption{Rejection power of GraXe in scintillation mode.} \label{tab:RFsci}
\end{table}
%%%%%

The efficiency of the event selection cuts and the achieved background rejection are evaluated over large samples of simulated data. Table \ref{tab:RFsci} summarizes the results. The first row indicates the size of the initial sample. The first cut rejects all events with energy depositions in the \emph{outer volume} (the space outside the graphene balloon). Further rejection is achieved by imposing that only one cluster is observed inside the ID. Finally, one imposes that the candidates are in the energy window around \Qbb. These cuts leave 67\% of the signal and suppress the intrinsic \bbtnu\ background by more than 5 orders of magnitude. 

To translate rejection power to real number of background events per year, we consider that the activity of a QUPID is 0.5 mBq of \BI\ and 0.5 mBq of \TL. With $\sim2\,800$ of these devices, it translates into $4.4 \times 10^7$ decays per year of \BI\ and about the same of \TL. From the suppression values in table \ref{tab:RFsci}, we can calculate that the number of \TL\ events passing the cuts (in the 1 FWHM region) is
\begin{equation}
6 \times 10^{-8} \cdot 4.4 \times 10^7 = 2.6\ ,
\end{equation}
and the number of \BI\ events is
\begin{equation}
5 \times 10^{-9} \cdot 4.4 \times 10^7 = 0.2\ .
\end{equation}

We also need to consider the intrinsic background, the standard \bbtnu\ decay. For a lifetime of $2.11 \times 10^{21}$ years \cite{Ackerman:2011gz}, this process would contribute with almost 34 events to the background
count, thus dominating (and considerably spoiling) the background rate. However, unlike the \BI\ and \TL\ backgrounds, in which the distribution of events in the energy window is essentially flat, the \bbtnu\ events are more probable in the left half. If we choose only the right side, the number of \bbtnu\ events per year decreases to 2.3, at the expense, of course, of reducing the signal efficiency by a factor of two.  

In summary, GraXe in scintillation mode would have a background rate (per unit of \bb\ isotope mass, energy and time) of:
\begin{equation}
b \simeq \frac{4}{1000 \cdot 123} = 3 \times 10^{-5}\ \ckky\ ,
\end{equation}
which is about one order of magnitude better than the background rate expected by the most competitive \bbonu\ experiments currently being constructed or commissioned \cite{GomezCadenas:2011it}. 

Let us now estimate the sensitivity of GraXe to the effective Majorana mass, \mbb. Following the method described in reference \cite{GomezCadenas:2010gs}, only three experimental parameters are needed: the energy resolution, the background rate and the detection efficiency. The first two have been discussed above. For the efficiency, a 35\% has been assumed, resulting from the product of the event selection efficiency (67\%, see table \ref{tab:RFsci}) and the energy cut to half of the window (50\%). The result is shown in figure \ref{fig:sensi} (blue, solid curve).

\section{Measuring ionization in Graxe} \label{sec:ion}
%%%
The inner volume of GraXe can be converted into a diode ionization chamber \cite{AprileBook}, such as the detector described in \cite{Giomataris:2008ap}, adding in the center a spherical electrode, that we call \emph{snitch} \cite{snitch}. Then, since graphene is metallic, a potential difference could be established between balloon and snitch, allowing the collection of ionization charges. The snitch would be an ultra-pure copper sphere of 1 cm radius (and thus $\sim38$ g of mass). Its specific activity would be very small, about 0.5 $\mu$Bq. The voltage would be set by means of a coaxial, ultra-pure copper rod connected to the snitch. This rod must be surrounded by another conductor set at a different potential to correct field distortions, as described in reference \cite{Giomataris:2008ap}. These two pieces would add only a few $\mu$Bq to the radioactive budget of the detector.

\begin{figure}[t!b!]
\begin{center}
\includegraphics[height=8cm]{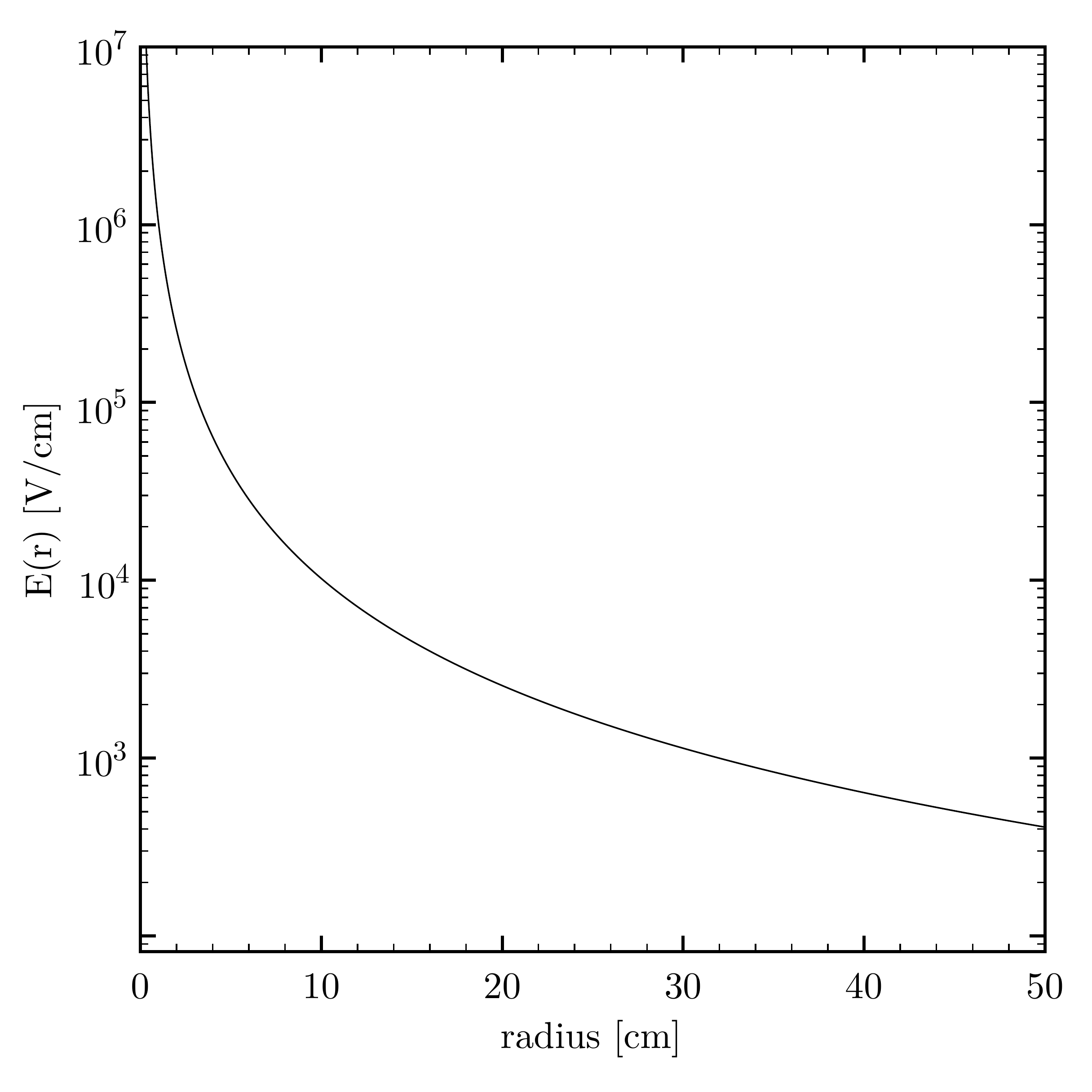}
\end{center}
\caption{The radial electric field in the inner volume of GraXe.} \label{fig.er}
\end{figure}

The electric field in the inner volume as a function of radius $r$ would be approximately:
\begin{equation}
E(r) = \frac{V}{1/r_a - 1/r_c} \ \frac{1}{r^2},
\end{equation}
where $r_{c}$ is the radius of the cathode (the graphene balloon), $r_{a}$ is the radius of the snitch, and $V$ is the potential difference between the electrodes. Figure \ref{fig.er} shows the electric field in the ID when we set $r_a=1$ cm, $r_c=44$ cm and $V=1$ MV. With such a configuration, the field is about 1 kV/cm near the graphene balloon, high enough to ensure the drift of the ionization charges. Since LXe is an excellent insulator, it appears possible to set the snitch at the large voltage required. The signal induced by the drifting electrons in the anode (the snitch) could be picked up by a low-noise charge amplifier. The effect of positive ions in the induction signal should be negligible \cite{AprileBook}. Alternatively, GraXe could operate as a gridded ionization chamber adding a Frisch grid \cite{Aprile:2009dv, AprileBook} close to the snitch. 

%%%%%
%%%%%

The measurement of the ionization charge allows to operate GraXe as a LXe time projection chamber. The initial time of the event, $t_0$, is given by the scintillation pulse, which also provides a measurement of the energy of the event, and locates the event vertex with a precision of the order of a few cm. The arrival of the ionization charge to the snitch allows, on the other hand, a much more precise measurement ($\sim 1$ mm) of the event in the radial coordinate. Furthermore, the measurement of the ionization in the snitch provides a second estimator of the event energy. Since scintillation and ionization are complementary and anti-correlated \cite{Aprile:2009dv}, their sum allows to infer a better energy resolution, perhaps as good as 4\% FWHM at \Qbb\ \cite{Conti:2003av}.

%%%%%%%%%%%%%%%%%%%%%%%%%%%%%%%%%%%%%%%%%%%%%%%%%%%%%%%%%%%%
\subsection{Sensitivity of GraXe using scintillation and ionization} \label{subsec:SensiIoni}

When using the ionization and scintillation mode combined, GraXe achieves good vertex resolution in the IV, of the order of a few millimeters in the radial coordinate, and an acceptable energy resolution, around 4\% FWHM at \Qbb. The improved vertex resolution can be used to better reject \Bi\ and \Tl\ events, which are often Compton events with an energy deposition separated by a few mm from the main energy cluster. Also, the improved energy resolution allows a much better rejection of the dominant \bbtnu\ background. 

%%%%%
\begin{table}[t!]
\begin{center}
\begin{tabular}{lcccc}
\toprule
Cuts & \bbonu\ & \bbtnu\ & \BI & \TL \\ \midrule
Initial sample  & $1.0\times10^5$ & $1.1 \times 10^9$ & $10^9$ & $10^8$ \\
No energy in OV & $9. 6 \times 10^4$ & $1.9 \times 10^6$ & 52 & 50 \\ 
Only one cluster & $7.7 \times 10^4$ & $1.6 \times 10^6$ & 4 & 4 \\ 
ROI (1 FWHM) & $5.8 \times 10^4$ & $1.1 \times 10^2$ & 1 & 2 \\ \midrule
%
%half side ROI  & $2.7 \times 10^4$ & $3.0 $ & $0.0 $ & $2.0$\\ \hline
%
Rejection factor & $0.58$ & $1.0 \times 10^{-7}$ & $1.0 \times 10^{-9}$ & $2.0\times 10^{-8}$ \\
\bottomrule
\end{tabular}
\end{center}
\caption{Rejection power of GraXe when measuring both the ionization and the scintillation.} \label{tab:RFion}
\end{table}%
%%%%%

Table \ref{tab:RFion} summarizes the background rejection and its effect on the signal. The cuts leave 58\% of the signal and suppress the \bbtnu\ background in the full ROI to negligible levels. Recalling the discussion of section \ref{subsec:sensi_sci}, we can calculate the background rate achieved in this operation mode. The number of \Tl\ events passing the cuts (in a 1-FWHM region) is
\begin{equation}
2 \times 10^{-8} \cdot 4.4 \times 10^7 = 0.9\, , \nonumber
\end{equation}
while the contribution of \BI\ is smaller than 0.1 events. Therefore:
\begin{equation}
b \simeq \frac{1}{1000 \times 98} = 1.0 \times 10^{-5}\ \ckky\, .
\end{equation}
This background rate is still two orders of magnitude worse than the irreducible background associated to the elastic electron scattering of solar neutrinos, calculated to be $1.65\times10^{-7}\ \ckky$ for \XE\ \cite{deBarros:2011qq}, leaving room for improvement. At these levels, neutron-induced backgrounds, not considered previously, might become a concern. For instance, $^{137}$Xe, a short-lived isotope resulting from the activation of xenon by neutrons, decays $\beta$ with a high $Q$-value. Nevertheless, the neutron flux can be easily suppressed by adding neutron shielding material surrounding the detector \cite{Carson:2004cb}, resulting in background rates below $10^{-7}$ \ckky\ \cite{Granena:2009it}.

The sensitivity to \mbb\ has been calculated as well for this operation mode, and it is shown in figure \ref{fig:sensi} (red, dashed curve). Notice that the improvement is not only due to the lower background rate, but also to the higher signal efficiency (the selection cut to half of the energy window is no longer needed, thanks to the improved energy resolution).

\section{Conclusions and outlook} \label{sec:map}
\begin{figure}[t!]
\centering
\includegraphics[height=10cm]{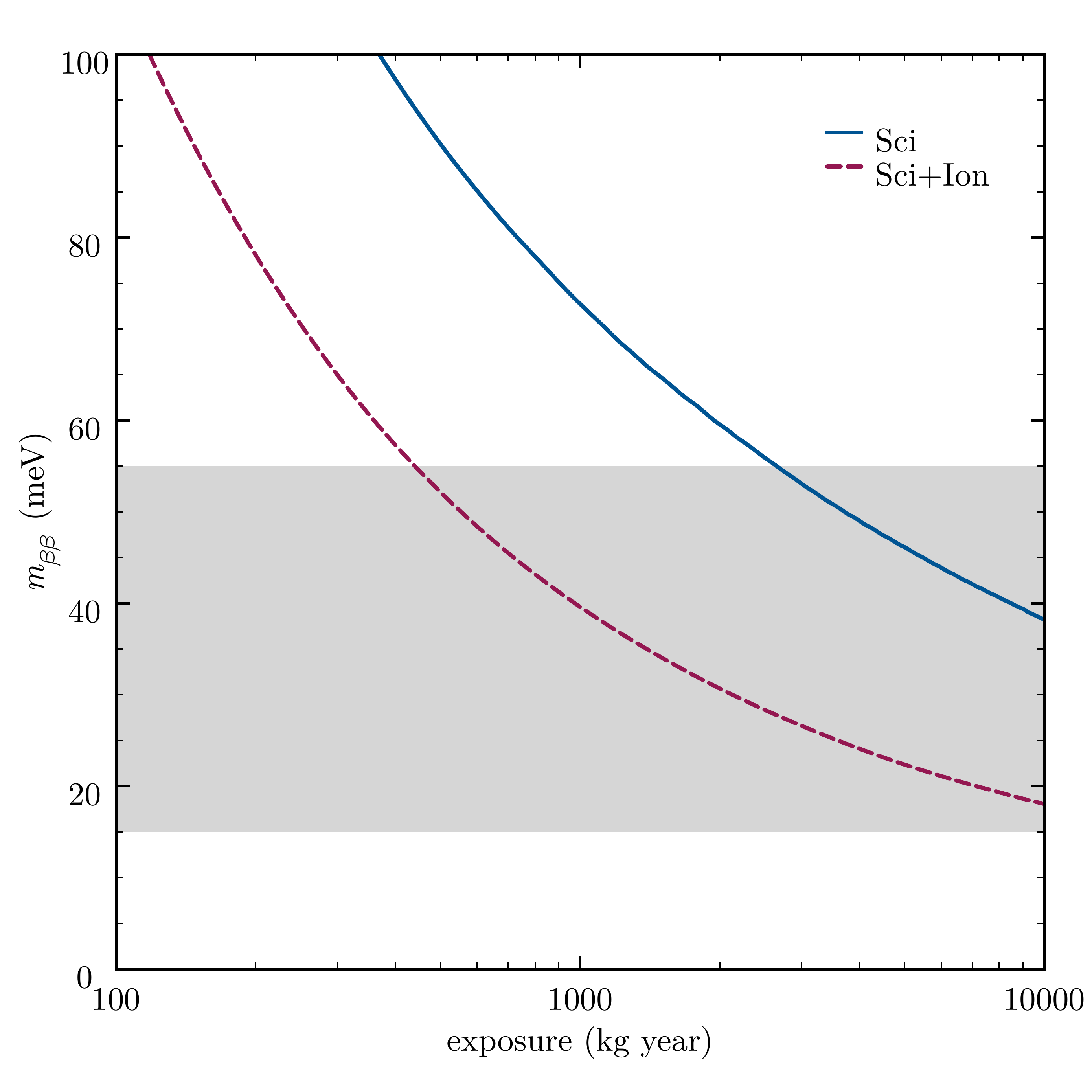}
\caption{Sensitivity (at 90\% CL) of GraXe to \mbb\ as a function of the exposure (isotope mass times data-taking time) for the two operation modes described in the text. The shaded band marks the inverted hierarchy of neutrino masses.} \label{fig:sensi}
\end{figure}

We have proposed a new experimental approach to search for neutrinoless double beta decay, the GraXe concept. We have shown that the combination of properties of xenon and graphene allows the construction of an extremely sensitive detector, able to fully explore the inverse hierarchy (see figure \ref{fig:sensi}).

%%%%%
%%%%%

All components of GraXe are standard technology save for the graphene balloon. Therefore, the obvious milestone towards a realistic detector would be the construction of a prototype of the balloon. A second milestone would be to test the whole GraXe concept at small scale. We suggest that some of the existing \emph{small} LXe detectors (such as XMASS phase I) could be used for this purpose. The final detector could be thought of as the inner part of a future large-scale, LXe, dark matter detector, such as XMASS, XENON \cite{Aprile:2011zz} or LUX \cite{McKinsey:2010zz}.

A similar vision to the one presented here (although the emphasis was placed on dark matter searches) was proposed by Arisaka and collaborators, the XAX detector \cite{Arisaka:2008mb}. A significant difference, though, was that they proposed the used of an acrylic balloon, rather than a graphene balloon. The use of an acrylic ballon, on the other hand, has two major disadvantages: acrylic is not transparent to the xenon VUV light, and therefore the balloon should be painted with a WLS, such as TPB; xenon diffuses through the acrylic surface, at an non-negligible rate. As a consequence the XAX concept would be very difficult to implement in a pure LXe detector (where it is essential that the enriched and the depleted xenon never mix), or in a LXe-LAr detector (one should reclaim the diffused xenon).

%%%%%%%%%%%%%%%%%%%%%%%%%%%%%%%%%%%%%%%%%%%%%%%%%%%%%%%%%%%%

\acknowledgments
The authors acknowledge support by the Spanish Ministerio de Ciencia e Innovaci\'on under grants CONSOLIDER-Ingenio 2010 CSD2008-0037 (CUP) and FPA2009-13697-C04-04.

%%%%%%%%%%%%%%%%%%%%%%%%%%%%%%%%%%%%%%%%%%%%%%%%%%%%%%%%%%%%

\bibliography{references}
\bibliographystyle{JHEP}

\end{document}